# Essential guidelines for computational method benchmarking

Lukas M. Weber [1,2]
Wouter Saelens [3,4]
Robrecht Cannoodt [3,4]
Charlotte Soneson [1,2] †
Alexander Hapfelmeier [5]
Paul P. Gardner [6]
Anne-Laure Boulesteix [7]
Yvan Saeys [3,4] *
Mark D. Robinson [1,2] *

[1] Institute of Molecular Life Sciences, University of Zurich, Zurich, Switzerland
[2] SIB Swiss Institute of Bioinformatics, University of Zurich, Zurich, Switzerland
[3] Data Mining and Modelling for Biomedicine, VIB Center for Inflammation Research, Ghent, Belgium
[4] Department of Applied Mathematics, Computer Science and Statistics, Ghent University, Ghent, Belgium
[5] Institute of Medical Informatics, Statistics and Epidemiology, Technical University of Munich, Munich, Germany
[6] Department of Biochemistry, University of Otago, Dunedin, New Zealand
[7] Institute for Medical Information Processing, Biometry and Epidemiology, Ludwig-Maximilians-University, Munich, Germany

† Current address: Friedrich Miescher Institute for Biomedical Research and SIB Swiss Institute of Bioinformatics, Basel, Switzerland

* Corresponding authors

## Abstract

In computational biology and other sciences, researchers are frequently faced with a choice between several computational methods for performing data analyses. Benchmarking studies aim to rigorously compare the performance of different methods using well-characterized benchmark datasets, to determine the strengths of each method or to provide recommendations regarding suitable choices of methods for an analysis. However, benchmarking studies must be carefully designed and implemented to provide accurate, unbiased, and informative results. Here, we summarize key practical guidelines and recommendations for performing high-quality benchmarking analyses, based on our experiences in computational biology.

## Introduction

Many fields of computational research are characterized by a growing number of available methods for data analysis. For example, at the time of writing, almost 400 methods are available for analyzing data from single-cell RNA-sequencing experiments [1]. For experimental researchers and method users, this represents both an opportunity and a challenge, since method choice can significantly affect conclusions.

Benchmarking studies are carried out by computational researchers to compare the performance of different methods, using reference datasets and a range of evaluation criteria. Benchmarks may be performed by authors of new methods to demonstrate performance improvements or other advantages; by independent groups interested in systematically comparing existing methods; or organized as community challenges. 'Neutral' benchmarking studies, i.e., those performed independently from new method development by authors without any perceived bias, and with a focus on the comparison itself, are especially valuable for the research community [2,3].

From our experience conducting benchmarking studies in computational biology, we have learned several key lessons that we aim to synthesize in this review. A number of previous reviews have addressed this topic from a range of perspectives, including: overall commentaries and recommendations on benchmarking design [2,4–9]; surveys of design practices followed by existing benchmarks [7]; the importance of neutral benchmarking studies [3]; principles for the design of real-data benchmarking studies [10,11] and simulation studies [12]; the incorporation of meta-analysis techniques into benchmarking [13–16]; the organization and role of community challenges [17,18]; and discussions on benchmarking design for specific types of methods [19,20]. More generally, benchmarking may be viewed as a form of meta-research [21].

Our aim is to complement previous reviews by providing a summary of essential guidelines for designing, performing, and interpreting benchmarks. While all guidelines are essential for a truly excellent benchmark, some are more fundamental than others. Our target audience consists of computational researchers who are interested in performing a benchmarking study, or who have already begun one. Our review spans the full 'pipeline' of benchmarking, from defining the scope to best practices for reproducibility. This includes crucial questions regarding design and evaluation principles: e.g., using rankings according to evaluation metrics to identify a set of high-performing methods, and then highlighting different strengths and tradeoffs among these.

The review is structured as a series of guidelines (Fig. 1), each explained in detail in the following sections. We use examples from computational biology; however, we expect that most arguments apply equally to other fields. We hope that these guidelines will continue the discussion on benchmarking design, as well as assisting computational researchers to design and implement rigorous, informative, and unbiased benchmarking analyses.

> **Fig. 1: Summary of guidelines**
>
> The guidelines in this review can be summarized in the following set of recommendations. Each recommendation is discussed in more detail in the corresponding section in the text.
>
> 1. Define the purpose and scope of the benchmark.
> 2. Include all relevant methods.
> 3. Select (or design) representative datasets.
> 4. Choose appropriate parameter values and software versions.
> 5. Evaluate methods according to key quantitative performance metrics.
> 6. Evaluate secondary measures including computational requirements, user-friendliness, installation procedures, and documentation quality.
> 7. Interpret results and provide recommendations from both user and method developer perspectives.
> 8. Publish results in an accessible format.
> 9. Design the benchmark to enable future extensions.
> 10. Follow reproducible research best practices, by making code and data publicly available.

## Defining the purpose and scope

The purpose and scope of a benchmark should be clearly defined at the beginning of the study, and will fundamentally guide the design and implementation. In general, we can define three broad types of benchmarking studies: (i) those by method developers, to demonstrate the merits of their approach (e.g., [22–26]); (ii) neutral studies performed to systematically compare methods for a certain analysis, either conducted directly by an independent group (e.g., [27–38]) or in collaboration with method authors (e.g., [39]); or, (iii) those organized in the form of a community challenge, such as those from the DREAM [40–44], FlowCAP [45,46], CASP [47,48], CAMI [49], Assemblathon [50,51], MAQC/SEQC [52–54], and GA4GH [55] consortia.

A neutral benchmark or community challenge should be as comprehensive as possible; although for any benchmark, there will be tradeoffs in terms of available resources. To minimize perceived bias, a research group conducting a neutral benchmark should be approximately equally familiar with all included methods, reflecting typical usage of the methods by independent researchers [3]. Alternatively, the group could include the original method authors, so that each method is evaluated under optimal conditions; methods whose authors decline to take part should be reported. In either case, bias due to focusing attention on particular methods should be avoided; e.g., when tuning parameters or fixing bugs. Strategies to avoid these types of biases, such as the use of blinding, have been previously proposed [10].

By contrast, when introducing a new method, the focus of the benchmark will be on evaluating the relative merits of the new method. This may be sufficiently achieved with a less extensive benchmark, e.g., by comparing against a smaller set of state-of-the-art and baseline methods. However, the benchmark must still be carefully designed to avoid disadvantaging any methods. For example, extensively tuning parameters for the new method while using default parameters for competing methods would result in a biased representation. Some advantages of a new method may fall outside the scope of a benchmark, e.g., a new method may enable more flexible analyses than previous methods (e.g., beyond two-group comparisons in differential analyses [22]).

Finally, results should be summarized in the context of the original purpose of the benchmark. A neutral benchmark or community challenge should provide clear guidelines for method users, and highlight weaknesses in current methods so that these can be addressed by method developers. On the other hand, benchmarks performed to introduce a new method should discuss what the new method offers compared to the current state-of-the-art, such as discoveries that would otherwise not be possible.

## Selection of methods

The selection of methods to include in the benchmark will be guided by the purpose and scope of the study. A neutral benchmark should include all available methods for a certain type of analysis. In this case, the publication describing the benchmark will also function as a review of the literature; a summary table describing the methods is a key output (e.g., see Fig. 2 in [27] or Table 1 in [31]). Alternatively, it may make sense to include only a subset of methods, by defining inclusion criteria: e.g., all methods that: (i) provide freely available software implementations; (ii) are available for commonly used operating systems; and, (iii) can successfully be installed without errors following a reasonable amount of trouble-shooting. Such criteria should be chosen without favoring any methods, and exclusion of any widely used methods should be justified. A useful strategy can be to involve method authors within the process, since they may provide additional details on optimal usage. In addition, community involvement can lead to new collaborations and inspire future method development. However, the overall neutrality and balance of the resulting research team should be maintained. Finally, if the benchmark is organized as a community challenge, the selection of methods will be determined by the participants. In this case, it is important to communicate the initiative widely; e.g., through an established network such as DREAM challenges. However, some authors may choose not to participate; a summary table documenting non-included methods should be provided in this case.

When developing a new method, it is generally sufficient to select a representative subset of existing methods to compare against. For example, this could consist of the current best-performing methods (if known), a simple 'baseline' method, and any methods that are widely used. The selection of competing methods should ensure an accurate and unbiased assessment of the relative merits of the new approach, compared to the current state-of-the-art. In fast-moving fields, for a truly excellent benchmark, method developers should be prepared to update their benchmarks or design them to easily allow extensions as new methods emerge.

## Selection (or design) of datasets

The selection of reference datasets is a critical design choice. If suitable publicly accessible datasets cannot be found, they will need to be generated or constructed, either experimentally or by simulation. Including a variety of datasets ensures that methods can be evaluated under a wide range of conditions. In general, reference datasets can be grouped into two main categories: simulated (or synthetic) and real (or experimental).

Simulated data have the advantage that a known true signal (or 'ground truth') can easily be introduced; e.g., whether a gene is differentially expressed. Quantitative performance metrics measuring the ability to recover the known truth can then be calculated. However, it is important to demonstrate that simulations accurately reflect relevant properties of real data, by inspecting empirical summaries of both simulated and real datasets (e.g., using automated tools [56]). The set of empirical summaries to use is context-specific; e.g., for single-cell RNA-sequencing, dropout profiles and dispersion-mean relationships should be compared [29]; for DNA methylation, correlation patterns among neighboring CpG sites should be investigated [57]; for comparing mapping algorithms, error profiles of the sequencing platforms should be considered [58]. Simplified simulations can also be useful, to evaluate a new method under a basic scenario, or to systematically test aspects such as scalability and stability. However, overly simplistic simulations should be avoided, since these will not provide useful information on performance. A further advantage of simulated data is that it is possible to generate as much data as required; e.g., to study variability and draw statistically valid conclusions.

Experimental data often do not contain a ground truth, making it difficult to calculate performance metrics. Instead, methods may be evaluated by comparing them against each other (e.g., overlap between sets of detected differential features [23]), or against a current widely accepted method or 'gold standard' (e.g., manual gating to define cell populations in high-dimensional cytometry [31,45], or fluorescence in situ hybridization to validate absolute copy number predictions [6]). In the context of supervised learning, the response variable to be predicted is known in the manually labeled training and test data. However, individual datasets should not be over-used, and using the same dataset for both method development and evaluation should be avoided, due to the risk of overfitting and overly optimistic results [59,60]. In some cases, it is also possible to design experimental datasets containing a ground truth. Examples include: (i) 'spiking in' synthetic RNA molecules at known relative concentrations [61] in RNA-sequencing experiments (e.g., [54,62]); (ii) large-scale validation of gene expression measurements by quantitative polymerase chain reaction (e.g., [54]); (iii) using genes located on sex chromosomes as a proxy for silencing of DNA methylation status (e.g., [26,63]); (iv) using fluorescence-activated cell sorting to sort cells into known subpopulations prior to single-cell RNA-sequencing (e.g., [29,64,65]); or, (v) mixing different cell lines to create 'pseudo-cells' [66]. However, it may be difficult to ensure that the ground truth represents an appropriate level of variability; e.g., the variability of spiked-in material, or whether method performance on cell line data is relevant to outbred populations. Alternatively, experimental datasets may be evaluated qualitatively, for example by judging whether each method can recover previous discoveries, although this strategy relies on the validity of previous results.

A further technique is to design 'semi-simulated' datasets that combine real experimental data with an 'in silico' (i.e., computational) spike-in signal, e.g., by combining cells or genes from 'null' (e.g., healthy) samples with a subset of cells or genes from samples expected to contain a true differential signal (examples include [22,67,68]). This strategy can create datasets with more realistic levels of variability and correlation, together with a ground truth.

Overall, there is no perfect reference dataset, and the selection of appropriate datasets will involve tradeoffs, e.g., regarding the level of complexity. Both simulated and experimental data should not be too 'simple' (e.g., two of the datasets in the FlowCAP-II challenge [45] gave perfect performance for several algorithms) or too 'difficult' (e.g., for the third dataset in FlowCAP-II, no algorithms performed well); in these situations, it can be impossible to distinguish performance. In some cases, individual datasets have also been found to be unrepresentative, leading to over-optimistic or otherwise biased assessment of methods (e.g., [69]). Overall, the key to truly excellent benchmarking is diversity of evaluations, i.e., using a range of metrics and datasets that span the range of those that might be encountered in practice, so that performance estimates can be credibly extrapolated.

## Parameters and software versions

Parameter settings can have a crucial impact on performance. Some methods have a large number of parameters, and tuning parameters to optimal values can require significant effort and expertise. For a neutral benchmark, a range of parameter values should ideally be considered for each method; although tradeoffs need to be considered regarding available time and computational resources. Importantly, the selection of parameter values should comply with the neutrality principle, i.e., certain methods should not be favored over others through more extensive parameter tuning.

There are three major strategies for choosing parameters. The first (and simplest) is to use default values for all parameters. Default parameters may be adequate for many methods, although this is difficult to judge in advance. While this strategy may be viewed as too simplistic for some neutral benchmarks, it reflects typical usage. We used default parameters in several neutral benchmarks where we were interested in performance for untrained users [27,70,71]. In addition, for [27], due to the large number of methods and datasets, total runtime was already around a week using 192 processor cores, necessitating judgment in the scope of parameter tuning. The second strategy is to choose parameters based on previous experience, or published values. This relies on familiarity with the methods and the literature, reflecting usage by expert users. The third strategy is to use a systematic or automated parameter tuning procedure; e.g., a 'grid search' across ranges of values for multiple parameters, or techniques such as cross-validation (e.g., [30]). The strategies may also be combined; e.g., setting non-critical parameters to default values, and performing a grid search for key parameters. Regardless, neutrality should be maintained: comparing methods with the same strategy makes sense, while comparing one method with default parameters against another with extensive tuning makes for an unfair comparison.

For benchmarks performed to introduce a new method, comparing against a single set of optimal parameter values for competing methods is often sufficient; these values may be selected during initial exploratory work, or by consulting documentation. However, as outlined above, bias may be introduced by tuning the parameters of the new method more extensively. The parameter selection strategy should be transparently discussed during the interpretation of the results, to avoid the risk of over-optimistic reporting due to expending more 'researcher degrees of freedom' on the new method [5,72].

Software versions can also influence results, especially if updates include major changes to methodology (e.g., [73]). Final results should generally be based on the latest available versions, which may require re-running some methods if updates become available during the course of a benchmark.

## Evaluation criteria: key quantitative performance metrics

Evaluation of methods will rely on one or more quantitative performance metrics (Fig. 2a). The choice of metric depends on the type of method and data. For example, for classification tasks with a ground truth, metrics include the true positive rate (TPR; sensitivity or recall), false positive rate (FPR; 1 - specificity), and false discovery rate (FDR). For clustering tasks, common metrics include the F1 score, adjusted Rand index, normalized mutual information, precision, and recall; some of these can be calculated at the cluster level as well as averaged (and optionally weighted) across clusters (e.g., these metrics were used to evaluate clustering methods in our own work [28,31] and by others [33,45,74]). Several of these metrics can also be compared visually to capture the tradeoff between sensitivity and specificity, e.g., using receiver operating characteristic (ROC) curves (TPR vs. FPR), TPR vs. FDR curves, or precision-recall (PR) curves (Fig. 2b). For imbalanced datasets, PR curves have been shown to be more informative than ROC curves [75,76]. These visual metrics can also be summarized as a single number, such as area under the ROC or PR curve; examples from our work include [22,29]. In addition to the tradeoff between sensitivity and specificity, a method's 'operating point' is important; in particular, whether the threshold used (e.g., 5% FDR) is calibrated to achieve the specified error rate. We often overlay this onto TPR-FDR curves by filled or open circles (e.g., Fig. 2b, generated using the iCOBRA package [77]); examples from our work include [22,23,25,78].

For methods with continuous-valued output (e.g., effect sizes or abundance estimates), metrics include the root mean square error, distance measures, Pearson correlation, sum of absolute log-ratios, log-modulus, and cross-entropy. As above, the choice of metric depends on the type of method and data (e.g., [41,79] used correlation, while [48] used root mean square deviation). Further classes of methods include those generating graphs, phylogenetic trees, overlapping clusters, or distributions; these require more complex metrics. In some cases, custom metrics may need to be developed (e.g., we defined new metrics for topologies of developmental trajectories in [27]). When designing custom metrics, it is important to assess their reliability across a range of prediction values (e.g., [80,81]). For some metrics, it may also be useful to assess uncertainty, e.g., via confidence intervals. In the context of supervised learning, classification or prediction accuracy can be evaluated by cross-validation, bootstrapping, or on a separate test dataset (e.g., [13,46]). In this case, procedures to split data into training and test sets should be appropriate for the data structure and the prediction task at hand (e.g., leaving out whole samples or chromosomes [82]).

Additional metrics that do not rely on a ground truth include measures of stability, stochasticity, and robustness. These measures may be quantified by running methods multiple times using different inputs or subsampled data (e.g., we observed substantial variability in performance for some methods in [29,31]). 'Missing values' may occur if a method does not return any values for a certain metric, e.g., due to a failure to converge or other computational issues such as excessive runtime or memory requirements (e.g., [27,29,31]). Fallback solutions such as imputation may be considered in this case [83], although these should be transparently reported. For non-deterministic methods (e.g., with random starts or stochastic optimization), variability in performance when using different random seeds or subsampled data should be characterized. Null comparisons can be constructed by randomizing group labels such that datasets do not contain any true signal, which can provide information on error rates (e.g., [22,25,26]). However, these must be designed carefully to avoid confounding by batch or population structure, and to avoid strong within-group batch effects that are not accounted for.

For most benchmarks, multiple metrics will be relevant. Focusing on a single metric can give an incomplete view: methods may not be directly comparable if they are designed for different tasks, and different users may be interested in different aspects of performance. Therefore, a crucial design decision is whether to focus on an overall ranking, e.g., by combining or weighting multiple metrics. In general, it is unlikely that a single method will perform best across all metrics, and performance differences between top-ranked methods for individual metrics can be small. Therefore, a good strategy is to use rankings from multiple metrics to identify a set of consistently high-performing methods, and then highlight the different strengths of these methods. For example, in [31], we identified methods that gave good clustering performance, and then highlighted differences in runtimes among these. In several studies, we have presented results in the form of a graphical summary of performance according to multiple criteria (examples include Fig. 3 in [27] and Fig. 5 in [29] from our work; and Fig. 2 in [39] and Fig. 6 in [32] from other authors). Identifying methods that consistently underperform can also be useful, to allow readers to avoid these.

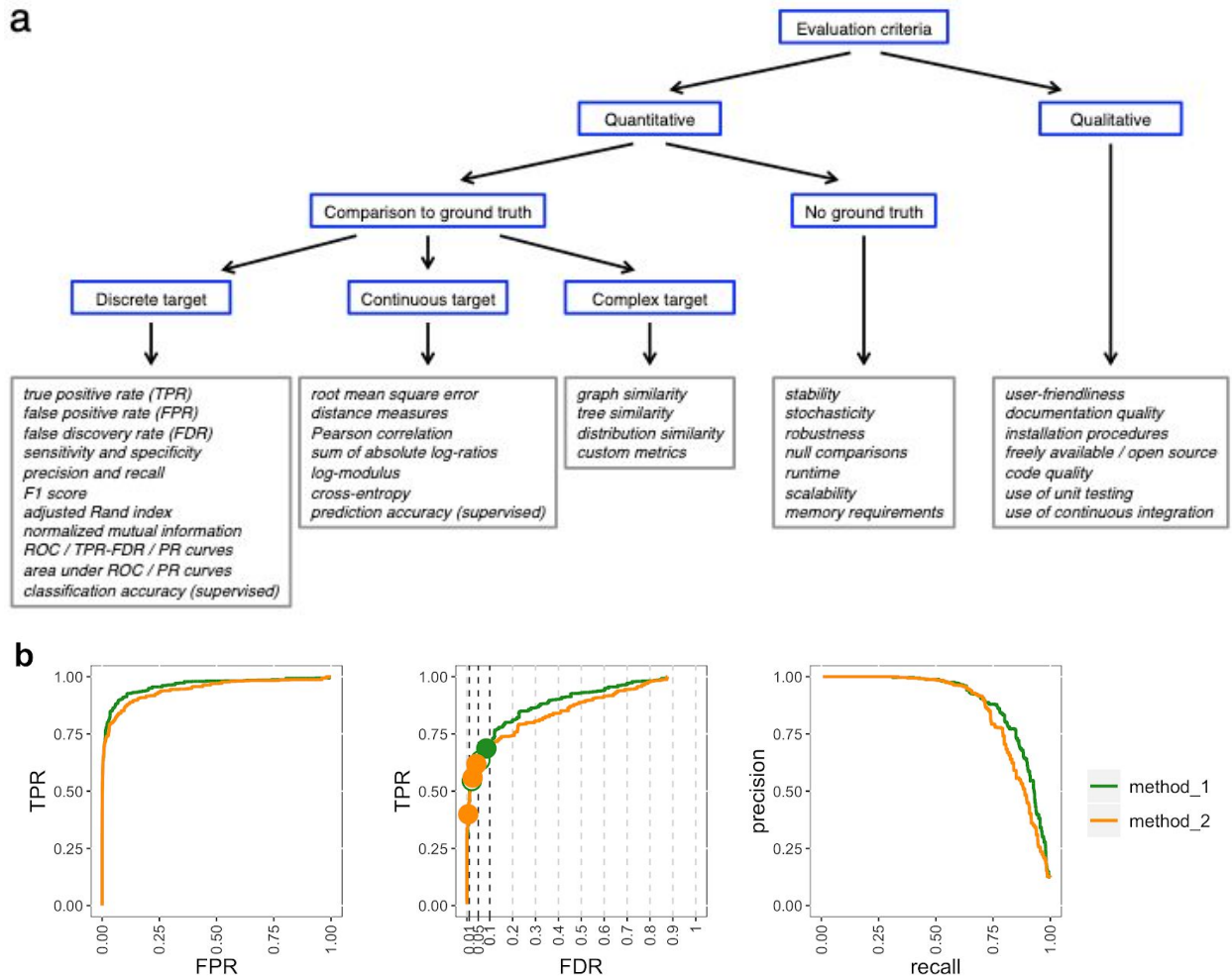

**Fig. 2.** Summary and examples of performance metrics. (a) Schematic overview of classes of frequently used performance metrics, including examples (gray boxes). (b) Examples of popular visualizations of quantitative performance metrics for classification methods, using reference datasets with a ground truth: (i) ROC curves, (ii) TPR vs. FDR curves (circles represent observed TPR and FDR at typical FDR thresholds of 1%, 5%, and 10%; filled circles indicate observed FDR lower than or equal to the imposed threshold), and (iii) PR curves. Visualizations in (b) generated using iCOBRA R/Bioconductor package [77].

## Evaluation criteria: secondary measures

In addition to the key quantitative performance metrics, methods should also be evaluated according to secondary measures including runtime, scalability, and other computational requirements, as well as qualitative aspects such as user-friendliness, installation procedures, code quality, and documentation quality (Fig. 2a). From the user perspective, the final choice of method may involve tradeoffs according to these measures: an adequately performing method may be preferable to a top-performing method that is especially difficult to use.

In our experience, runtimes and scalability can vary enormously between methods (e.g., in our work, runtimes for cytometry clustering algorithms [31] and metagenome analysis tools [79] ranged across multiple orders of magnitude for the same datasets). Similarly, memory and other computational requirements can vary widely. Runtimes and scalability may be investigated systematically, e.g., by varying the number of cells or genes in a single-cell RNA-sequencing dataset [28,29]. In many cases, there is a tradeoff between performance and computational requirements. In practice, if computational requirements for a top-performing method are prohibitive, then a different method may be preferred by some users.

User-friendliness, installation procedures, and documentation quality can also be highly variable [84,85]. Streamlined installation procedures can be ensured by distributing the method via standard package repositories, such as CRAN and Bioconductor for R, or PyPI for Python. Alternative options include GitHub and other code repositories or institutional websites; however, these options do not provide users with the same guarantees regarding reliability and documentation quality. Availability across multiple operating systems and within popular programming languages for data analysis are also important. Availability of graphical user interfaces can further extend accessibility, although graphical-only methods hinder reproducibility and are thus difficult to include in a systematic benchmark.

For many users, freely available and open source software will be preferred, since it is more broadly accessible and can be adapted by experienced users. From the developer perspective, code quality and use of software development best practices, such as unit testing and continuous integration, are also important. Similarly, adherence to commonly used data formats (e.g., GFF/GTF files for genomic features, BAM/SAM files for sequence alignment data, or FCS files for flow or mass cytometry data) greatly improves accessibility and extensibility.

High-quality documentation is critical, including help pages and tutorials. Ideally, all code examples in the documentation should be continually tested; e.g., as Bioconductor does, or through continuous integration.

## Interpretation, guidelines, and recommendations

For a truly excellent benchmark, results must be clearly interpreted from the perspective of the intended audience. For method users, results should be summarized in the form of recommendations. An overall ranking of methods (or separate rankings for multiple evaluation criteria) can provide a useful overview. However, as mentioned above, some methods may not be directly comparable (e.g., since they are designed for different tasks), and different users may be interested in different aspects of performance. In addition, it is unlikely that there will be a clear 'winner' across all criteria, and performance differences between top-ranked methods can be small. Therefore, an informative strategy is to use the rankings to identify a set of high-performing methods, and to highlight the different strengths and tradeoffs among these methods. The interpretation may also involve biological or other domain knowledge to establish the scientific relevance of differences in performance. Importantly, neutrality principles should be preserved during the interpretation.

For method developers, the conclusions may include guidelines for possible future development of methods. By assisting method developers to focus their research efforts, high-quality benchmarks can have significant impact on the progress of methodological research.

Limitations of the benchmark should be transparently discussed. For example, in [27] we used default parameters for all methods, while in [31] our datasets relied on manually gated reference cell populations as the ground truth. Without a thorough discussion of limitations, a benchmark runs the risk of misleading readers; in extreme cases, this may even harm the broader research field by guiding research efforts in the wrong directions.

## Publication and reporting of results

The publication and reporting strategy should emphasize clarity and accessibility. Visualizations summarizing multiple performance metrics can be highly informative for method users (examples include Fig. 3 in [27] and Fig. 5 in [29] from our own work; as well as Fig. 6 in [32]). Summary tables are also useful as a reference (e.g., [31,45]). Additional visualizations, such as flow charts to guide the choice of method for different analyses, are a helpful way to engage the reader (e.g., Fig. 5 in [27]).

For extensive benchmarks, online resources enable readers to interactively explore the results (examples from our work include [27,29], which allow users to filter metrics and datasets). Fig. 3 displays an example of an interactive website from one of our benchmarks [27], which facilitates exploration of results and assists users with choosing a suitable method. While tradeoffs should be considered in terms of the amount of work required, these efforts are likely to have significant benefit for the community.

In most cases, results will be published in a peer-reviewed article. For a neutral benchmark, the benchmark will be the main focus of the paper. For a benchmark to introduce a new method, the results will form one part of the exposition. We highly recommend publishing a preprint prior to peer review (e.g., on bioRxiv or arXiv) to speed up distribution of results, broaden accessibility, and solicit additional feedback. In particular, direct consultation with method authors can generate highly useful feedback (examples from our work are described in the acknowledgments in [79,86]). Finally, at publication time, considering open access options will further broaden accessibility.

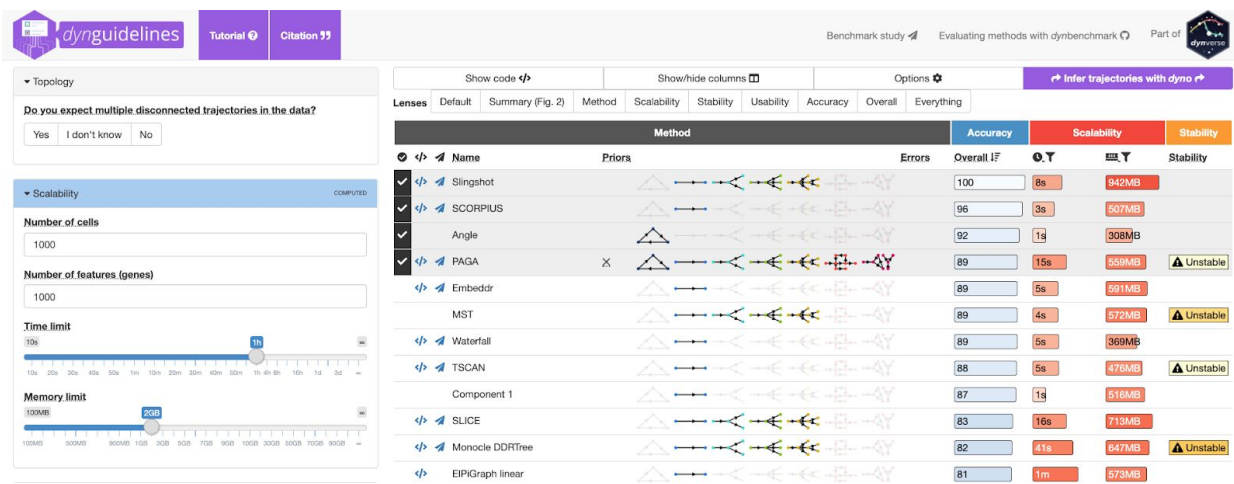

**Fig. 3.** Example of an interactive website allowing users to explore the results of one of our benchmarking studies [27]. This website was created using the Shiny framework in R.

## Enabling future extensions

Since new methods are continually emerging [1], benchmarks can quickly become out of date. To avoid this, a truly excellent benchmark should be extensible. For example, creating public repositories containing code and data allows other researchers to build on the results to include new methods or datasets, or to try different parameter settings or pre-processing procedures (examples from our work include [27–31]). In addition to raw data and code, it is useful to distribute pre-processed and/or results data (examples include [28,29,77] from our work and [74,87,88] from others), especially for computationally intensive benchmarks. This may be combined with an interactive website, where users can upload results from a new method, to be included in an updated comparison either automatically or by the original authors (e.g., [35,89,90]). 'Continuous' benchmarks, which are continually updated, are especially convenient (e.g., [91]), but may require significant additional effort.

## Reproducible research best practices

Reproducibility of research findings has become an increasing concern in numerous areas of study [92]. In computational sciences, reproducibility of code and data analyses has been recognized as a useful 'minimum standard' that enables other researchers to verify analyses [93]. Access to code and data have previously enabled method developers to uncover potential errors in published benchmarks due to suboptimal usage of methods [73,94,95]. Journal publication policies can play a crucial role in encouraging authors to follow these practices [96]; experience shows that statements that code and data are 'available on request' are often insufficient [97]. In the context of benchmarking, code and data availability also provides further benefits: for method users, code repositories serve as a source of annotated code to run methods and build analysis pipelines, while for developers, code repositories can act as a prototype for future method development work.

Parameter values (including random seeds) and software versions should be clearly reported, to ensure complete reproducibility. For methods that are run using scripts, these will be recorded within the scripts. In R, the command 'sessionInfo()' gives a complete summary of package versions, the version of R, and the operating system. For methods only available via graphical interfaces, parameters and versions must be recorded manually. Reproducible workflow frameworks, such as the Galaxy platform [98], can also be helpful. A summary table or spreadsheet of parameter values and software versions can be published as supplementary information along with the publication describing the benchmark (e.g., Supporting Information Table S1 in our study [31]).

Automated workflow management tools and specialized tools for organizing benchmarks provide sophisticated options for setting up benchmarks and creating a reproducible record, including software environments, package versions, and parameter values. Examples include SummarizedBenchmark [99], DataPackageR [100], workflowr [101], and Dynamic Statistical Comparisons [102]. Some tools (e.g., workflowr) also provide streamlined options for publishing results online. In machine learning, OpenML provides a platform to organize and share benchmarks [103]. More general tools for managing computational workflows, including Snakemake [104], Make, Bioconda [105], and conda, can be customized to capture setup information. Containerization tools such as Docker and Singularity may be used to encapsulate a software environment for each method, preserving the package version as well as dependency packages and the operating system, and facilitating distribution of methods to end users (e.g., in our study [27]). Best practices from software development are also useful, including unit testing and continuous integration.

Many free online resources are available for sharing code and data, including GitHub and Bitbucket, repositories for specific data types (e.g., ArrayExpress [106], the Gene Expression Omnibus [107], and FlowRepository [108]), and more general data repositories (e.g., figshare, Dryad, Zenodo, Bioconductor ExperimentHub, and Mendeley Data). Customized resources (examples from our work include [29,77]) can be designed when additional flexibility is needed. Several repositories allow the creation of 'digital object identifiers' (DOIs) for code or data objects. In general, preference should be given to publicly funded repositories, which provide greater guarantees for long-term archival stability [84,85].

An extensive literature exists on best practices for reproducible computational research (e.g., [109]). Some practices (e.g., containerization) may involve significant additional work; however, in our experience, almost all efforts in this area prove useful, especially by facilitating later extensions by ourselves or other researchers.

| Principle (see Fig. 1) | How essential? | Tradeoffs | Potential pitfalls |
|---|---|---|---|
| 1. Defining the purpose and scope | ∗ ∗ ∗ | How comprehensive the benchmark should be | - Scope too broad: too much work given available resources<br>- Scope too narrow: unrepresentative and possibly misleading results |
| 2. Selection of methods | ∗ ∗ ∗ | Number of methods to include | - Excluding key methods |
| 3. Selection (or design) of datasets | ∗ ∗ ∗ | Number and types of datasets to include | - Subjectivity in the choice of datasets: e.g., selecting datasets that are unrepresentative of real-world applications<br>- Too few datasets or simulation scenarios<br>- Overly simplistic simulations |
| 4. Parameter and software versions | ∗ ∗ | Amount of parameter tuning | - Extensive parameter tuning for some methods while using default parameters for others (e.g., competing methods) |
| 5. Evaluation criteria: key quantitative performance metrics | ∗ ∗ ∗ | Number and types of performance metrics | - Subjectivity in the choice of metrics: e.g., selecting metrics that do not translate to real-world performance<br>- Metrics that give over-optimistic estimates of performance<br>- Methods may not be directly comparable according to individual metrics (e.g., if methods are designed for different tasks) |
| 6. Evaluation criteria: secondary measures | ∗ ∗ | Number and types of performance metrics | - Subjectivity of qualitative measures such as user-friendliness, installation procedures, and documentation quality<br>- Subjectivity in relative weighting between multiple metrics<br>- Measures such as runtime and scalability depend on processor speed and memory |
| 7. Interpretation, guidelines, and recommendations | ∗ ∗ | Generality vs. specificity of recommendations | - Performance differences between top-ranked methods may be minor<br>- Different readers may be interested in different aspects of performance |
| 8. Publication and reporting of results | ∗ | Amount of resources to dedicate to building online resources | - Online resources may not be accessible (or may no longer run) several years later |
| 9. Enabling future extensions | ∗ ∗ | Amount of resources to dedicate to ensuring extensibility | - Selection of methods or datasets for future extensions may be unrepresentative (e.g., due to requests from method authors) |
| 10. Reproducible research best practices | ∗ ∗ | Amount of resources to dedicate to reproducibility | - Some tools may not be compatible or accessible several years later |

**Table 1.** Summary of our views regarding 'how essential' each principle is for a truly excellent benchmark (higher number of stars means more central to the evaluation); along with examples of key tradeoffs and potential pitfalls relating to each principle.

## Discussion

In this review, we have described a set of key principles for designing a high-quality computational benchmark. In our view, elements of all of these principles are essential. However, we have also emphasized that any benchmark will involve tradeoffs, due to limited expertise and resources, and that some principles are less central to the evaluation. Table 1 provides a summary of examples of key tradeoffs and pitfalls related to benchmarking, along with our judgment of how truly 'essential' each principle is.

There are a number of potential pitfalls that may arise from benchmarking studies (Table 1). For example, subjectivity in the choice of datasets or evaluation metrics could bias the results. In particular, a benchmark that relies on unrepresentative data or metrics that do not translate to real-world scenarios may be misleading by showing poor performance for methods that otherwise perform well. This could harm method users, who may select an inappropriate method for their analyses, as well as method developers, who may be discouraged from pursuing promising methodological approaches. In extreme cases, this could negatively affect the research field by influencing the direction of research efforts. A thorough discussion of the limitations of a benchmark can help avoid these issues. Over the longer term, critical evaluations of published benchmarks, so-called meta-benchmarks, will also be informative [10,13,14].

Well-designed benchmarking studies provide highly valuable information for users and developers of computational methods, but require careful consideration of a number of important design principles. In this review, we have discussed a series of guidelines for rigorous benchmarking design and implementation, based on our experiences in computational biology. We hope these guidelines will assist computational researchers to design high-quality, informative benchmarks, which will contribute to scientific advances through informed selection of methods by users and targeting of research efforts by developers.

## Author contributions

LMW proposed the project and drafted the manuscript. WS, RC, CS, AH, PPG, ALB, YS, and MDR contributed ideas and references and contributed to drafting of the manuscript. YS and MDR supervised the project. All authors read and approved the final manuscript.

## Acknowledgments

The authors thank members of the Robinson Lab at the University of Zurich and Saeys Lab at Ghent University for valuable feedback.

## Funding

We acknowledge funding support from the Swiss National Science Foundation (grant numbers 310030_175841 and CRSII5_177208). In addition, this project has been made possible in part (grant number 2018-182828) by the Chan Zuckerberg Initiative DAF, an advised fund of the Silicon Valley Community Foundation. LMW was supported by a Forschungskredit (Candoc) grant from the University of Zurich (FK-17-100). WS and RC are supported by the Fonds Wetenschappelijk Onderzoek. YS is an ISAC Marylou Ingram Scholar. ALB was supported by individual grants BO3139/2-3 and BO3139/4-3 from the German Research Foundation (DFG) and by the German Federal Ministry of Education and Research under grant number 01IS18036A (MCML).

## Competing interests

The authors declare that they have no competing interests.

# References

1. Zappia L, Phipson B, Oshlack A. Exploring the single-cell RNA-seq analysis landscape with the scRNA-tools database. PLoS Computational Biology. 2018;14: e1006245.

2. Boulesteix A-L, Binder H, Abrahamowicz M, Sauerbrei W. On the necessity and design of studies comparing statistical methods. Biometrical Journal. 2018;60: 216–218.

3. Boulesteix A-L, Lauer S, Eugster MJA. A plea for neutral comparison studies in computational sciences. PLoS ONE. 2013;8: e61562.

4. Peters B, Brenner SE, Wang E, Slonim D, Kann MG. Putting benchmarks in their rightful place: The heart of computational biology. PLoS Computational Biology. 2018;14: e1006494.

5. Boulesteix A-L. Ten simple rules for reducing overoptimistic reporting in methodological computational research. PLoS Computational Biology. 2015;11: e1004191.

6. Zheng S. Benchmarking: contexts and details matter. Genome Biology. 2017;18.

7. Mangul S, Martin LS, Hill BL, Lam AK-M, Distler MG, Zelikovsky A, et al. Systematic benchmarking of omics computational tools. Nature Communications. 2019;10: 1393.

8. Norel R, Rice JJ, Stolovitzky G. The self-assessment trap: can we all be better than average? Molecular Systems Biology. 2011;7: 537.

9. Aniba MR, Poch O, Thompson JD. Issues in bioinformatics benchmarking: the case study of multiple sequence alignment. Nucleic Acids Research. 2010;38: 7353–7363.

10. Boulesteix A-L, Wilson R, Hapfelmeier A. Towards evidence-based computational statistics: lessons from clinical research on the role and design of real-data benchmark studies. BMC Medical Research Methodology. 2017;17.

11. Boulesteix A-L, Hable R, Lauer S, Eugster MJA. A statistical framework for hypothesis testing in real data comparison studies. The American Statistician. 2015;69: 201–212.

12. Morris TP, White IR, Crowther MJ. Using simulation studies to evaluate statistical methods. Statistics in Medicine. 2019;38: 2074–2102.

13. Gardner PP, Watson RJ, Morgan XC, Draper JL, Finn RD, Morales SE, et al. Identifying accurate metagenome and amplicon software via a meta-analysis of sequence to taxonomy benchmarking studies. PeerJ. 2019;7: e6160.

14. Gardner PP, Paterson JM, Ashari-Ghomi F, Umu SU, McGimpsey S, Pawlik A. A meta-analysis of bioinformatics software benchmarks reveals that publication-bias unduly influences software accuracy. bioRxiv. 2017;092205.

15. Evangelou E, Ioannidis JPA. Meta-analysis methods for genome-wide association studies and beyond. Nature Reviews Genetics. 2013;14: 379–389.

16. Hong F, Breitling R. A comparison of meta-analysis methods for detecting differentially expressed genes in microarray experiments. Bioinformatics. 2008;24: 374–382.

17. Boutros PC, Margolin AA, Stuart JM, Califano A, Stolovitzky G. Toward better benchmarking:

challenge-based methods assessment in cancer genomics. Genome Biology. 2014;15: 462.

18. Friedberg I, Wass MN, Mooney SD, Radivojac P. Ten simple rules for a community computational challenge. PLoS Computational Biology. 2015;11: e1004150.

19. Van Mechelen I, Boulesteix A-L, Dangl R, Dean N, Guyon I, Hennig C, et al. Benchmarking in cluster analysis: A white paper. arXiv. 2018;1809.10496.

20. Angers-Loustau A, Petrillo M, Bengtsson-Palme J, Berendonk T, Blais B, Chan K-G, et al. The challenges of designing a benchmark strategy for bioinformatics pipelines in the identification of antimicrobial resistance determinants using next generation sequencing technologies. F1000Research. 2018;7: 459.

21. Ioannidis JPA. Meta-research: Why research on research matters. PLoS Biology. 2018;16: e2005468.

22. Weber LM, Nowicka M, Soneson C, Robinson MD. diffcyt: Differential discovery in high-dimensional cytometry via high-resolution clustering. Communications Biology. 2019;2: 183.

23. Nowicka M, Robinson MD. DRIMSeq: a Dirichlet-multinomial framework for multivariate count outcomes in genomics. F1000Research. 2016;5: 1356.

24. Levine JH, Simonds EF, Bendall SC, Davis KL, Amir E-AD, Tadmor MD, et al. Data-driven phenotypic dissection of AML reveals progenitor-like cells that correlate with prognosis. Cell. 2015;162: 184–197.

25. Zhou X, Lindsay H, Robinson MD. Robustly detecting differential expression in RNA sequencing data using observation weights. Nucleic Acids Research. 2014;42: e91.

26. Law CW, Chen Y, Shi W, Smyth GK. voom: precision weights unlock linear model analysis tools for RNA-seq read counts. Genome Biology. 2014;15: R29.

27. Saelens W, Cannoodt R, Todorov H, Saeys Y. A comparison of single-cell trajectory inference methods. Nature Biotechnology. 2019;

28. Duò A, Robinson MD, Soneson C. A systematic performance evaluation of clustering methods for single-cell RNA-seq data. F1000Research. 2018;7: 1141.

29. Soneson C, Robinson MD. Bias, robustness and scalability in single-cell differential expression analysis. Nature Methods. 2018;15: 255–261.

30. Saelens W, Cannoodt R, Saeys Y. A comprehensive evaluation of module detection methods for gene expression data. Nature Communications. 2018;9: 1090.

31. Weber LM, Robinson MD. Comparison of clustering methods for high-dimensional single-cell flow and mass cytometry data. Cytometry Part A. 2016;89A: 1084–1096.

32. Korthauer K, Kimes PK, Duvallet C, Reyes A, Subramanian A, Teng M, et al. A practical guide to methods controlling false discovery rates. Genome Biology. 2019;

33. Freytag S, Tian L, Lönnstedt I, Ng M, Bahlo M. Comparison of clustering tools in R for medium-sized 10x Genomics single-cell RNA-sequencing data. F1000Research. 2018;7: 1297.

34. Baruzzo G, Hayer KE, Kim EJ, Di Camillo B, FitzGerald GA, Grant GR. Simulation-based comprehensive benchmarking of RNA-seq aligners. Nature Methods. 2017;14: 135–139.

35. Kanitz A, Gypas F, Gruber AJ, Gruber AR, Martin G, Zavolan M. Comparative assessment of

methods for the computational inference of transcript isoform abundance from RNA-seq data. Genome Biology. 2015;16.

36. Soneson C, Delorenzi M. A comparison of methods for differential expression analysis of RNA-seq data. BMC Bioinformatics. 2013;14.

37. Rapaport F, Khanin R, Liang Y, Pirun M, Krek A, Zumbo P, et al. Comprehensive evaluation of differential gene expression analysis methods for RNA-seq data. Genome Biology. 2013;14.

38. Dillies M-A, Rau A, Aubert J, Hennequet-Antier C, Jeanmougin M, Servant N, et al. A comprehensive evaluation of normalization methods for Illumina high-throughput RNA sequencing data analysis. Briefings in Bioinformatics. 2012;14: 671–683.

39. Sage D, Kirshner H, Pengo T, Stuurman N, Min J, Manley S, et al. Quantitative evaluation of software packages for single-molecule localization microscopy. Nature Methods. 2015;12: 717–724.

40. Weirauch MT, Cote A, Norel R, Annala M, Zhao Y, Riley TR, et al. Evaluation of methods for modeling transcription factor sequence specificity. Nature Biotechnology. 2013;31: 126–134.

41. Costello JC, Heiser LM, Georgii E, Gönen M, Menden MP, Wang NJ, et al. A community effort to assess and improve drug sensitivity prediction algorithms. Nature Biotechnology. 2014;32: 1202–1212.

42. Küffner R, Zach N, Norel R, Hawe J, Schoenfeld D, Wang L, et al. Crowdsourced analysis of clinical trial data to predict amyotrophic lateral sclerosis progression. Nature Biotechnology. 2015;33: 51–57.

43. Ewing AD, Houlahan KE, Hu Y, Ellrott K, Caloian C, Yamaguchi TN, et al. Combining tumor genome simulation with crowdsourcing to benchmark somatic single-nucleotide-variant detection. Nature Methods. 2015;12: 623–630.

44. Hill SM, Heiser LM, Cokelaer T, Unger M, Nesser NK, Carlin DE, et al. Inferring causal molecular networks: empirical assessment through a community-based effort. Nature Methods. 2016;13: 310–318.

45. Aghaeepour N, Finak G, The FlowCAP Consortium, The DREAM Consortium, Hoos H, Mosmann TR, et al. Critical assessment of automated flow cytometry data analysis techniques. Nature Methods. 2013;10: 228–238.

46. Aghaeepour N, Chattopadhyay P, Chikina M, Dhaene T, Van Gassen S, Kursa M, et al. A benchmark for evaluation of algorithms for identification of cellular correlates of clinical outcomes. Cytometry Part A. 2016;89A: 16–21.

47. Moult J, Fidelis K, Kryshtafovych A, Schwede T, Tramontano A. Critical assessment of methods of protein structure prediction (CASP) — Round XII. Proteins. 2018;86: 7–15.

48. Moult J, Fidelis K, Kryshtafovych A, Schwede T, Tramontano A. Critical assessment of methods of protein structure prediction: Progress and new directions in round XI. Proteins. 2016;84: 4–14.

49. Sczyrba A, Hofmann P, Belmann P, Koslicki D, Janssen S, Dröge J, et al. Critical Assessment of Metagenome Interpretation—a benchmark of metagenomics software. Nature Methods. 2017;14: 1063–1071.

50. Earl D, Bradnam K, St John J, Darling A, Lin D, Fass J, et al. Assemblathon 1: A competitive assessment of de novo short read assembly methods. Genome Research. 2011;21: 2224–2241.

51. Bradnam KR, Fass JN, Alexandrov A, Baranay P, Bechner M, Birol I, et al. Assemblathon 2: evaluating de novo methods of genome assembly in three vertebrate species. GigaScience. 2013;2:

1–31.

52. MAQC Consortium. The MicroArray Quality Control (MAQC) project shows inter- and intraplatform reproducibility of gene expression measurements. Nature Biotechnology. 2006;24: 1151–1161.

53. MAQC Consortium. The Microarray Quality Control (MAQC)-II study of common practices for the development and validation of microarray-based predictive models. Nature Biotechnology. 2010;28: 827–838.

54. SEQC/MAQC-III Consortium. A comprehensive assessment of RNA-seq accuracy, reproducibility and information content by the Sequencing Quality Control Consortium. Nature Biotechnology. 2014;32: 903–914.

55. Krusche P, Trigg L, Boutros PC, Mason CE, De La Vega FM, Moore BL, et al. Best practices for benchmarking germline small-variant calls in human genomes. Nature Biotechnology. 2019;

56. Soneson C, Robinson MD. Towards unified quality verification of synthetic count data with countsimQC. Bioinformatics. 2017;34: 691–692.

57. Korthauer K, Chakraborty S, Benjamini Y, Irizarry RA. Detection and accurate false discovery rate control of differentially methylated regions from whole genome bisulfite sequencing. Biostatistics. 2018; 1–17.

58. Caboche S, Audebert C, Lemoine Y, Hot D. Comparison of mapping algorithms used in high-throughput sequencing: application to Ion Torrent data. BMC Genomics. 2014;15: 264.

59. Grimm DG, Azencott C-A, Aicheler F, Gieraths U, MacArthur DG, Samocha KE, et al. The evaluation of tools used to predict the impact of missense variants is hindered by two types of circularity. Human Mutation. 2015;36: 513–523.

60. Jelizarow M, Guillemot V, Tenenhaus A, Strimmer K, Boulesteix A-L. Over-optimism in bioinformatics: an illustration. Bioinformatics. 2010;26: 1990–1998.

61. Jiang L, Schlesinger F, Davis CA, Zhang Y, Li R, Salit M, et al. Synthetic spike-in standards for RNA-seq experiments. Genome Research. 2011;21: 1543–1551.

62. Garalde DR, Snell EA, Jachimowicz D, Sipos B, Lloyd JH, Bruce M, et al. Highly parallel direct RNA sequencing on an array of nanopores. Nature Methods. 2018;15: 201–206.

63. Fang F, Hodges E, Molaro A, Dean M, Hannon GJ, Smith AD. Genomic landscape of human allele-specific DNA methylation. Proceedings of the National Academy of Sciences of the USA. 2012;109: 7332–7337.

64. The Tabula Muris Consortium. Single-cell transcriptomics of 20 mouse organs creates a Tabula Muris. Nature. 2018;562: 367–372.

65. Zheng GXY, Terry JM, Belgrader P, Ryvkin P, Bent ZW, Wilson R, et al. Massively parallel digital transcriptional profiling of single cells. Nature Communications. 2017;8: 14049.

66. Tian L, Dong X, Freytag S, Lê Cao K-A, Su S, JalalAbadi A, et al. Benchmarking single cell RNA-sequencing analysis pipelines using mixture control experiments. Nature Methods. 2019;16: 479–487.

67. Arvaniti E, Claassen M. Sensitive detection of rare disease-associated cell subsets via representation learning. Nature Communications. 2017;8: 1–10.

68. Rigaill G, Balzergue S, Brunaud V, Blondet E, Rau A, Rogier O, et al. Synthetic data sets for the

69. Löwes B, Chauve C, Ponty Y, Giegerich R. The BRaliBase dent — a tale of benchmark design and interpretation. Briefings in Bioinformatics. 2017;18: 306–311.

70. Couronné R, Probst P, Boulesteix A-L. Random forest versus logistic regression: a large-scale benchmark experiment. BMC Bioinformatics. 2018;19: 270.

71. Schneider J, Hapfelmeier A, Thöres S, Obermeier A, Schulz C, Pförringer D, et al. Mortality Risk for Acute Cholangitis (MAC): a risk prediction model for in-hospital mortality in patients with acute cholangitis. BMC Gastroenterology. 2016;16: 15.

72. Hu Q, Greene CS. Parameter tuning is a key part of dimensionality reduction via deep variational autoencoders for single cell RNA transcriptomics. Pacific Symposium on Biocomputing. 2019;24: 362–373.

73. Vaquero-Garcia J, Norton S, Barash Y. LeafCutter vs. MAJIQ and comparing software in the fast moving field of genomics. bioRxiv. 2018;463927.

74. Wiwie C, Baumbach J, Röttger R. Comparing the performance of biomedical clustering methods. Nature Methods. 2015;12: 1033–1038.

75. Saito T, Rehmsmeier M. The precision-recall plot is more informative than the ROC plot when evaluating binary classifiers on imbalanced datasets. PLoS ONE. 2015;10: e0118432.

76. Powers DMW. Visualization of tradeoff in evaluation: from precision-recall & PN to LIFT, ROC & BIRD. arXiv. 2015;1505.00401.

77. Soneson C, Robinson MD. iCOBRA: open, reproducible, standardized and live method benchmarking. Nature Methods. 2016;13: 283.

78. Soneson C, Love MI, Robinson MD. Differential analyses for RNA-seq: transcript-level estimates improve gene-level inferences. F1000Research. 2016;4: 1521.

79. Lindgreen S, Adair KL, Gardner PP. An evaluation of the accuracy and speed of metagenome analysis tools. Scientific Reports. 2016;6: 19233.

80. Gurevich A, Saveliev V, Vyahhi N, Tesler G. QUAST: quality assessment tool for genome assemblies. Bioinformatics. 2013;29: 1072–1075.

81. Narzisi G, Mishra B. Comparing de novo genome assembly: the long and short of it. PLoS ONE. 2011;6: e19175.

82. Schreiber J, Singh R, Bilmes J, Noble WS. A pitfall for machine learning methods aiming to predict across cell types. bioRxiv. 2019;512434.

83. Bischl B, Schiffner J, Weihs C. Benchmarking local classification methods. Computational Statistics. 2013;28: 2599–2619.

84. Mangul S, Martin LS, Eskin E, Blekhman R. Improving the usability and archival stability of bioinformatics software. Genome Biology. 2019;20: 47.

85. Mangul S, Mosqueiro T, Abdill RJ, Duong D, Mitchell K, Sarwal V, et al. Challenges and recommendations to improve installability and archival stability of omics computational tools. bioRxiv. 2019;452532.

identification of key ingredients for RNA-seq differential analysis. Briefings in Bioinformatics. 2018;19: 65–76.

69. Löwes B, Chauve C, Ponty Y, Giegerich R. The BRaliBase dent — a tale of benchmark design and interpretation. Briefings in Bioinformatics. 2017;18: 306–311.

70. Couronné R, Probst P, Boulesteix A-L. Random forest versus logistic regression: a large-scale benchmark experiment. BMC Bioinformatics. 2018;19: 270.

71. Schneider J, Hapfelmeier A, Thöres S, Obermeier A, Schulz C, Pförringer D, et al. Mortality Risk for Acute Cholangitis (MAC): a risk prediction model for in-hospital mortality in patients with acute cholangitis. BMC Gastroenterology. 2016;16: 15.

72. Hu Q, Greene CS. Parameter tuning is a key part of dimensionality reduction via deep variational autoencoders for single cell RNA transcriptomics. Pacific Symposium on Biocomputing. 2019;24: 362–373.

73. Vaquero-Garcia J, Norton S, Barash Y. LeafCutter vs. MAJIQ and comparing software in the fast moving field of genomics. bioRxiv. 2018;463927.

74. Wiwie C, Baumbach J, Röttger R. Comparing the performance of biomedical clustering methods. Nature Methods. 2015;12: 1033–1038.

75. Saito T, Rehmsmeier M. The precision-recall plot is more informative than the ROC plot when evaluating binary classifiers on imbalanced datasets. PLoS ONE. 2015;10: e0118432.

76. Powers DMW. Visualization of tradeoff in evaluation: from precision-recall & PN to LIFT, ROC & BIRD. arXiv. 2015;1505.00401.

77. Soneson C, Robinson MD. iCOBRA: open, reproducible, standardized and live method benchmarking. Nature Methods. 2016;13: 283.

78. Soneson C, Love MI, Robinson MD. Differential analyses for RNA-seq: transcript-level estimates improve gene-level inferences. F1000Research. 2016;4: 1521.

79. Lindgreen S, Adair KL, Gardner PP. An evaluation of the accuracy and speed of metagenome analysis tools. Scientific Reports. 2016;6: 19233.

80. Gurevich A, Saveliev V, Vyahhi N, Tesler G. QUAST: quality assessment tool for genome assemblies. Bioinformatics. 2013;29: 1072–1075.

81. Narzisi G, Mishra B. Comparing de novo genome assembly: the long and short of it. PLoS ONE. 2011;6: e19175.

82. Schreiber J, Singh R, Bilmes J, Noble WS. A pitfall for machine learning methods aiming to predict across cell types. bioRxiv. 2019;512434.

83. Bischl B, Schiffner J, Weihs C. Benchmarking local classification methods. Computational Statistics. 2013;28: 2599–2619.

84. Mangul S, Martin LS, Eskin E, Blekhman R. Improving the usability and archival stability of bioinformatics software. Genome Biology. 2019;20: 47.

85. Mangul S, Mosqueiro T, Abdill RJ, Duong D, Mitchell K, Sarwal V, et al. Challenges and recommendations to improve installability and archival stability of omics computational tools. bioRxiv. 2019;452532.

86. Freyhult EK, Bollback JP, Gardner PP. Exploring genomic dark matter: a critical assessment of the performance of homology search methods on noncoding RNA. Genome Res. Cold Spring Harbor Lab; 2007;17: 117–125.

87. Bokulich NA, Rideout JR, Mercurio WG, Shiffer A, Wolfe B, Maurice CF, et al. mockrobiota: a public resource for microbiome bioinformatics benchmarking. mSystems. 2016;1: e00062–16.

88. Conchúir SO, Barlow KA, Pache RA, Ollikainen N, Kundert K, O'Meara MJ, et al. A web resource for standardized benchmark datasets, metrics, and Rosetta protocols for macromolecular modeling and design. PLoS ONE. 2015;10: e0130433.

89. Cope LM, Irizarry RA, Jaffee HA, Wu Z, Speed TP. A benchmark for Affymetrix GeneChip expression measures. Bioinformatics. 2004;20: 323–331.

90. Irizarry RA, Wu Z, Jaffee HA. Comparison of Affymetrix GeneChip expression measures. Bioinformatics. 2006;22: 789–794.

91. Barton M. nucleotid.es: an assembler catalogue. http://nucleotid.es/. 2014;

92. Ioannidis JPA. Why most published research findings are false. PLoS Medicine. 2005;2: e124.

93. Peng RD. Reproducible research in computational science. Science. 2011;334: 1226–1227.

94. Zhou X, Robinson MD. Do count-based differential expression methods perform poorly when genes are expressed in only one condition? Genome Biology. 2015;16: 222.

95. Zhou X, Oshlack A, Robinson MD. miRNA-Seq normalization comparisons need improvement. RNA. 2013;19: 733–734.

96. Hofner B, Schmid M, Edler L. Reproducible research in statistics: A review and guidelines for the Biometrical Journal. Biometrical Journal. 2016;58: 416–427.

97. Boulesteix A-L, Janitza S, Hornung R, Probst P, Busen H, Hapfelmeier A. Making complex prediction rules applicable for readers: Current practice in random forest literature and recommendations. Biometrical Journal. 2018;

98. Afgan E, Baker D, Batut B, Van Den Beek M, Bouvier D, Čech M, et al. The Galaxy platform for accessible, reproducible and collaborative biomedical analyses: 2018 update. Nucleic Acids Research. 2018;46: W537–W544.

99. Kimes PK, Reyes A. Reproducible and replicable comparisons using SummarizedBenchmark. Bioinformatics. 2018; 1–3.

100. Finak G, Mayer B, Fulp W, Obrecht P, Sato A, Chung E, et al. DataPackageR: Reproducible data preprocessing, standardization and sharing using R/Bioconductor for collaborative data analysis. Gates Open Research. 2018;2.

101. Blischak J, Carbonetto P, Stephens M. workflowr: organized + reproducible + shareable data science in R. https://jdblischak.github.io/workflowr/. 2018;

102. Wang G, Stephens M, Carbonetto P. DSC: Dynamic Statistical Comparisons. https://stephenslab.github.io/dsc-wiki/index.html. 2018;

103. Vanschoren J, van Rijn JN, Bischl B, Torgo L. OpenML: networked science in machine learning. SIGKDD Explorations. 2014;15: 49–60.

104. Köster J, Rahmann S. Snakemake — a scalable bioinformatics workflow engine. Bioinformatics.

2012;28: 2520–2522.

105.   Grüning B, Dale R, Sjödin A, Chapman BA, Rowe J, Tomkins-Tinch CH, et al. Bioconda: sustainable and comprehensive software distribution for the life sciences. Nature Methods. 2018;15: 475–476.

106.   Kolesnikov N, Hastings E, Keays M, Melnichuk O, Tang YA, Williams E, et al. ArrayExpress update — simplifying data submissions. Nucleic Acids Research. 2015;43: D1113–D1116.

107.   Barrett T, Wilhite SE, Ledoux P, Evangelista C, Kim IF, Tomashevsky M, et al. NCBI GEO: archive for functional genomics data sets - update. Nucleic Acids Research. 2013;41: D991–D995.

108.   Spidlen J, Breuer K, Rosenberg C, Kotecha N, Brinkman RR. FlowRepository: A resource of annotated flow cytometry datasets associated with peer-reviewed publications. Cytometry Part A. 2012;81A: 727–731.

109.   Sandve GK, Nekrutenko A, Taylor J, Hovig E. Ten simple rules for reproducible computational research. PLoS Computational Biology. 2013;9: e1003285.